\documentclass{article}
\newcommand{\bra}[1]{\langle #1 |}
\newcommand{\ket}[1]{| #1 \rangle}
\newcommand{\Bra}[1]{\bigl\langle #1 \bigr|}
\newcommand{\Ket}[1]{\bigl| #1 \bigr\rangle}
\newcommand{\brkt}[2]{\langle #1 | #2 \rangle}
\newcommand{\ve}[1]{\mathbf{#1}}
\newcommand{\op}[1]{\widehat{#1}}
\newcommand{\Eq}[1]{Eq.~(\ref{#1})}
\newcommand{\BO}[1]{\mathbin{\mathrm{#1}}}
\newcommand{\Id}{{\rm 1\!\!I}}
\newcommand{\MO}[1]{\mathord{\mathrm{#1}}}

\usepackage{cite}
\usepackage{amsmath,amssymb}
\usepackage{url}
\usepackage{graphicx}

\begin{document}
\title{Some Notes on Quantum Information Theory and 
Emerging Computing Technologies}

\author{Alexander Yu.~Vlasov}
\date{}

\maketitle

\begin{abstract}
It is considered an interdependence of the theory of quantum computing 
and some perspective information technologies. A couple of illustrative 
and useful examples are discussed. The reversible
computing from very beginning had the serious impact
on the design of quantum computers and it is revisited first. 
Some applications of ternary circuits are also quite instructive 
and it may be useful in the quantum information theory. 
\end{abstract}

\section{Introduction}
The theory of reversible computations produced important income to 
the quantum computing, because for an evolution of a quantum system
described by Schr\"odinger equation the time inversion $t \to -t$
is valid operation. 
So the understanding of possibility to implement an universal computer
using reversible devices\cite{Ben73} was important for the development
of first models of the quantum computing machines \cite{BenEr}.

The reversible computation is also actively studied because 
according to the Landauer principle \cite{Land} the heat generation may be 
reduced to an arbitrary small value only for the design with reversible,
``conservative'' logic gates. Yet, the Landauer limit $k T \ln 2$
for irreversible operation could be treated as an extremely small 50 
years ago at the time of publication of the mentioned paper \cite{Land}.

Even in 1984 in his report about the quantum-mechanical computers
Feynman \cite{FeyComp} noted that because the actual dissipation
is still much bigger (about $10^{10} kT$ for a transistor with 
about $10^{11}$ atoms at the moment of his talk) the discussion 
about such small quantities and logical elements with a few or 
single atom is rather ``ridiculous'' and
``[s]uch nonsense is very entertaining to professors like me.''
Nevertheless, nowadays logical gates with single atom, ion, electron or
photon are already standard subject for real experiments in
area of quantum computing and communications \cite{NCQC,qcroad} and 
due to some prognoses $kT$ limit may be actual for real processors
to next decade or so \cite[Fig.~17]{Key05}.

There were some discussions about the Landauer principle those above
the scope of this presentation, but anyway the research of reversible 
computations is in the state of quite active development during an enough
long time and it is producing a mutually advantageous connection with
the quantum information theory. 

Some applications of a reversible design in the quantum information theory 
are discussed in this presentation.  
Examples of binary and ternary circuits
together with a brief excursus to the many-valued logic are also provided.

\section{Quantum Computations and Reversibility}

There is very close relation between classical reversible computations
and quantum information theory, because any reversible classical function
directly corresponds to a quantum one \cite{FeyComp,QAlg}. If there is a 
discrete system with $N$ states, there are $N!$ reversible functions, corresponding
to permutations of these states. A permutation $S$ has a standard representation 
by $N \times N$ matrix $\op{S}$ with $N$ nonzero elements $\op{S}_{ij}=1$ for
$S: i \mapsto j$, i.e., $\op{S}_{ij} = \delta_{S(i)j}$. 

If $\ve{e}_i$ denotes an element of the basis of the $N$-dimensional vector space, 
then due to such a definition $\op{S} \ve{e}_i = \ve{e}_{S(i)}$ and it produces a 
standard linear representation of the permutation group. 
In the quantum information science the Dirac notation is often used for 
simplification: $\ket{i}$ --- are $N$ basic vectors
(instead of $\ve{e}_i$), $\ket{i}\bra{j}$ --- is the matrix with only nonzero element 
$\op{M}_{ij}=1$ and $\brkt{i}{j}=(\ve{e}_i,\ve{e}_j)=\delta_{ij}$ --- is the scalar 
product (for the basis it is the Kronecker delta).
In such a notation an equation
\begin{equation}
 \op{S} = \sum_i \Ket{S(i)}\Bra{i}, \quad \op{S}\ket{i} = \Ket{S(i)} 
\label{transp}
\end{equation}
is hold.

Roughly speaking, a model with reversible circuits may be ``translated'' 
into the language of the quantum information theory after 
a formal change of the notation $0, 1, \ldots$ to $\ket{0}, \ket{1}, \ldots$,
but it is useful also to remember about specific properties of quantum
systems. The qubit is a quantum analogue of bits, but besides
it may be in any superposition of the basic states, i.e.,
$\alpha \ket{0} + \beta \ket{1}$,  
$|\alpha|^2 + |\beta|^2 = 1$. The qutrit is a ternary analogue
of the qubit and most general state may be described as
$\alpha \ket{0} + \beta \ket{1} + \gamma \ket{2}$,  
$|\alpha|^2 + |\beta|^2 + |\gamma|^2 = 1$. 

\begin{figure}[!t]
\centering
\includegraphics[width=4cm]{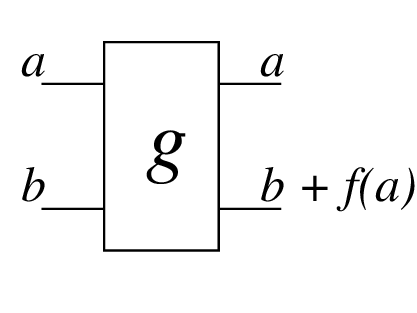}
\caption{Reversible gate for irreversible function}
\label{qfe}
\end{figure}

However, the problem of the realization of a classical algorithms on a quantum 
computer is also devoting an attention and it is actively discussed in the presented
work. Let us consider a question about the representation of an arbitrary irreversible
function or a ``gate'' using reversible one. 
Both in the quantum information theory and in the reversible computations
a method with an auxiliary system is widely used \cite{QAlg}. Instead of
the function (gate) $y = f(a)$ a gate with two ``wires'' is used 
\begin{equation}
g \colon (a,b) \mapsto \bigl(a,b+f(a)\bigr),
\label{revf}
\end{equation}
see Fig.~\ref{qfe}. 
Such a gate has inverse: $(a,b) \mapsto (a,b-f(a))$ and
for $b = 0$ reproduces initial function $(a,0) \mapsto (a,f(a))$.

Really, due to the fixed size of a number in the computer representation instead of
the addition in \Eq{revf} the modular arithmetic should be used, e.g.,
bitwise addition modulo 2 (so-called binary XOR, eXclusive OR operation) or
addition modulo $N_{\max}$ (some fixed value like $2^{32}$, $2^{64}$, etc.) may
be applied.

Using Dirac notation \Eq{revf} could be formally rewritten as
\begin{equation}
\op{G} \ket{x,z} = \Ket{x,z+f(x)}.
\label{qrevf}
\end{equation}
Formally, \Eq{qrevf} is a proper definition of $\op{G}$, because
for the definition of a linear operator it is enough
to describe transformations of all basic vectors. On the
other hand it is possible to produce ``more constructive''
description close to ideas of quantum control.

The CONTROLLED NOT gate may be considered as the first example of such approach.  
Already mentioned Feynman work about quantum-mechanical computers \cite{FeyComp} 
discussed that gate. It may be written for two bits as 
\[\MO{CNOT}\colon (a,b) \mapsto (a, b \BO{XOR} a).\]
It corresponds to \Eq{revf} for the trivial case of the identity
function $f(a)=a$, but it is useful anyway as a simplest 
example of a controlled gate and due to numerous applications.

The algorithm of control for CNOT may be represented as
\begin{verse}
{\bf if} $a$ {\bf then} NOT $b$ {\bf else} $b$
\end{verse}
and there is an instructive method to represent such operations
in the quantum computation. It is {\em conditional quantum dynamics}
\cite{condyn}. Using Dirac notation adopted in \cite{condyn}
the CNOT gate may be written as
\begin{equation}
  \op{\MO{CNOT}} = \ket{0}\bra{0} \otimes  \op{\Id} + 
 \ket{1}\bra{1} \otimes \op{\MO{NOT}},
\label{Cn}
\end{equation}
where $\Id$ is the identity operation, i.e., the unit matrix.

In the more general case there are more than two alternatives
and both $a$ and $b$ may represent more than two binary
values
\begin{verse}
{\bf case} $a$ {\bf of}\\
 \quad 0 : $F_0(b)$\\
 \quad 1 : $F_1(b)$\\
 \quad \ldots \\
 \quad k : $F_{\rm k}(b)$\\
{\bf end}
\end{verse}
and if instead of the reversible classical functions $F_{\rm k}$
to use unitary quantum operators $\op{U}_k$, then
the {\em conditional quantum dynamics} may be written as \cite{condyn}
\begin{equation}
  \op{U} = \ket{0}\bra{0} \otimes  \op{U}_0 + 
  \ket{1}\bra{1} \otimes  \op{U}_1 + \cdots + 
 \ket{k}\bra{k} \otimes \op{U}_k.
\label{Cu}
\end{equation}

The quantum notation used in \Eq{Cu} almost directly corresponds
to the ``{\bf case} control flow'' example above. The tensor product signs 
$\otimes$ are used for construction of states and operators
for composite systems in the quantum mechanics, i.e., notation like 
$\ket{00}$ or $\ket{0}\ket{0}$ could be rewritten in more pedantic
way as $\ket{0} \otimes \ket{0}$. For operators sign $\otimes$  
often should not be omitted to prevent confusion with usual multiplication 
(composition).

The terms $\op{P}_k=\ket{k}\bra{k}$ in \Eq{Cn} and \Eq{Cu} are 
{\em projectors}. We have $\op{P}_k\ket{k} = \ket{k}$ and 
$\op{P}_k\ket{j} = 0$, $j \ne k$ and so each term with 
$\op{P}_a$ selects only necessary states $\ket{a}\ket{b}$ and 
applies $\op{U}_a\ket{b}$ to the second variable.

Yet another example is the irreversible binary function AND.
Because an argument here is the {\em pair} of bits, a reversible  
analog is \[\mathrm{T} \colon (a_1,a_2,b) 
\mapsto \bigl(a_1,a_2,b \BO{XOR} (a_1 \BO{AND} a_2)\bigr).\]
The T is called CONTROLLED CONTROLLED NOT or
Toffoli gate \cite{FeyComp,tgt}. 

The Toffoli gate is important, because with CNOT gates it is
still not possible to represent any logical circuit, but
this problem may be resolved with Toffoli gates. So it is 
{\em universal} gate in the classical meaning \cite{FeyComp,tgt}.

It corresponds to a formal algorithm
\begin{verse}
{\bf if} ($a_1$ AND $a_2$) {\bf then} NOT $b$ {\bf else} $b$
\end{verse}
and in quantum notation it is
\[ \op{T} = \op{P}_1 \otimes \op{P}_1 \otimes \op{\MO{NOT}} +
 (\op{P}_0 \otimes \op{P}_0+\op{P}_0 \otimes \op{P}_1+
\op{P}_1 \otimes \op{P}_0) \otimes \op{\Id}. \]

So, two-bit reversible gates in the classical case are not enough to
create any function and it is necessary to use three-bit gates. 
However, in the quantum case two-bit gates may me used for
construction of any quantum circuit \cite{DBE95,Bar95,DiV95,SMB04}.

In the more general case \Eq{revf} may be applied to arbitrary Boolean
function with $N$ bits of input for $a$ and $M$ bits of output for $f(a)$. 
In such a case \Eq{revf} requires $M$ auxiliary bits of input with zeros $b = 0$
and produces $N$ bits of ``garbage'' due to the copying of an initial state $a$.

Formally, in the classical case any presentation of numbers (``radix'') may be 
used, e.g., binary, ternary, decimal numbers and so on. 
In the quantum information theory some representation may be preferable, 
e.g., the qubit is most appropriate in many cases, sometimes prime numbers 
are more convenient than factored ones, but it is above the scope of 
this work.

Computers repeat some elementary set of operations many times and
in each such step reversible circuits such as \Eq{revf} need for clean 
zero bits and generate new bits with garbage. 
It may be visualized using idea of some tapes with initially zero 
values in each cell and with ``garbage'' or ``history'' data those
are moving on each step of the computing device . Similar design was from very
beginning used in models of reversible and quantum computing machines 
\cite{Ben73,BenEr}.

\section{Qubits and Qutrits}

For Toffoli gate a cost of the reversibility in comparison with AND is
one extra zero bit and two bits of garbage. On the other hand it
is possible to use more ``economical'' design if to work with
nonbinary gates. For example, it is enough to let even
one wire to represent three values instead of two to include
the universal set of irreversible Boolean gates into a reversible 
system:
\begin{equation}
\begin{array}{|lc|l} 
 \BO{AND}_{23} & & \BO{OR}_{23} \\
\begin{array}{|cc|c|cc|}
 a & b &\rightarrow& a & b \\ \hline
 \mathbf 0 & \mathbf 0 & & \mathbf 0 & 0  \\
 \mathbf 0 & \mathbf 1 & & \mathbf 0 & 1  \\
 0 & 2 & & 1 & 2  \\
 \mathbf 1 & \mathbf 0 & & \mathbf 0 & 2  \\
 \mathbf 1 & \mathbf 1 & & \mathbf 1 & 0  \\
 1 & 2 & & 1 & 1  
\end{array}
& &
\begin{array}{|cc|c|cc|}
 a & b &\rightarrow& a & b \\ \hline
 \mathbf 0 & \mathbf 0 & & \mathbf 0 & 0  \\
 \mathbf 0 & \mathbf 1 & & \mathbf 1 & 1  \\
 0 & 2 & & 0 & 2  \\
 \mathbf 1 & \mathbf 0 & & \mathbf 1 & 2  \\
 \mathbf 1 & \mathbf 1 & & \mathbf 1 & 0  \\
 1 & 2 & & 0 & 1  
\end{array}
\end{array}
\label{twotri}
\end{equation}
The boldface numbers in \Eq{twotri} mark the inclusion of AND, OR
as subsets of suggested reversible operations. \Eq{twotri} may be represented
as compositions of two reversible steps. The first one for both
cases is the operation $b \mapsto b - a$ (modulo 3). The second step is
the application of (controlled) NOT gate to $a$ either for $b=2$ to
implement AND$_{23}$ or for $b=1$ to do OR$_{23}$.

Such implementation requires six states 
instead of eight (three bits) for Toffoli gate. Similar methods are known in 
reversible computation \cite{MVQCI,revsyn} and often for convenience and 
symmetry both values are ternary, yet for such a case formally there
are nine states without 
a self-evident advantage in comparison with Toffoli gate.

If both values are ternary, then instead of \Eq{twotri} with six possible
alternatives corresponding to an exchange of the values of $b$, there are 
much more (2160) variants of extensions of AND and OR gates. 
However, a reversible function with two values may be written
in form \(f\colon (a,b) \mapsto \bigl(f_1(a,b),f_2(a,b)\bigl)\),
there the second function $f_2$ is auxiliary. 

It may be shown that there are only ten alternative for the first function 
$f_1$ appropriate for representation of AND, OR operations. Between them
only two are symmetric and here is chosen one pair of such
functions:
\begin{equation}
\begin{array}{c|ccc}
\hline
\BO{AND}_3^\circ & 0 & 1 & 2 \\ \hline
0   & 0 & 0 & 2 \\
1   & 0 & 1 & 1 \\
2   & 2 & 1 & 2
\end{array}
\qquad
\begin{array}{c|ccc}
\hline
\BO{OR}_3^\circ & 0 & 1 & 2 \\ \hline
0  & 0 & 1 & 0 \\
1  & 1 & 1 & 2 \\
2  & 0 & 2 & 2
\end{array}
\label{cyclog}
\end{equation}

\begin{figure}[!t]
\centering
\includegraphics[width=4cm]{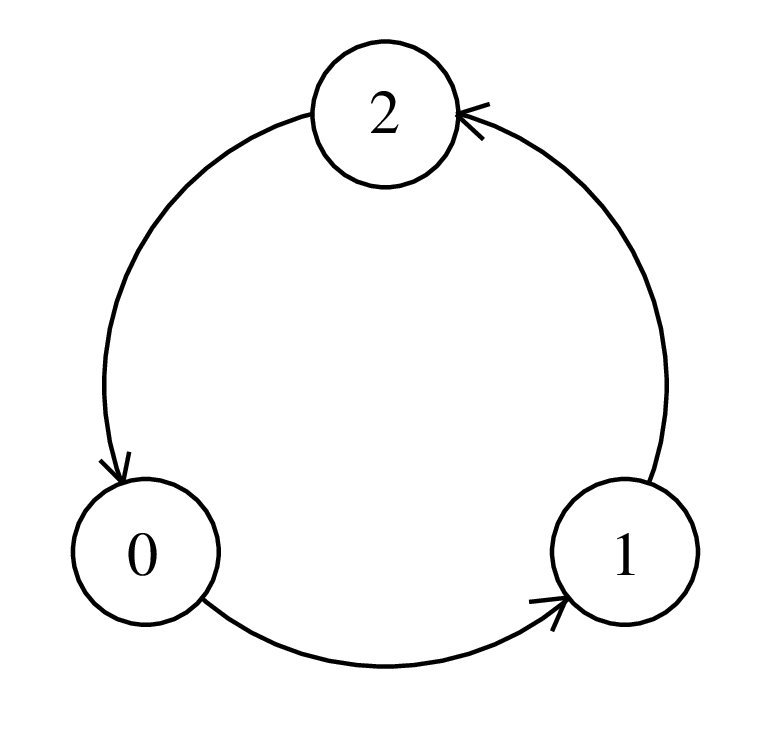}
\caption{Cyclic relation}
\label{c123}
\end{figure}

The operations may be described using an idea of the selection
``the previous or the same'' ($\preceq_3^\circ$) and 
``the next or the same'' ($\succeq_3^\circ$) between
two values with respect to a nontransitive connected relation 
$0 \prec_3^\circ 1$, $1 \prec_3^\circ 2$, $2 \prec_3^\circ 0$ 
depicted on Fig.~\ref{c123}. More precisely, 
$a \BO{AND}_3^\circ b$ returns $a$ if $a \prec_3^\circ b$ and $b$ 
otherwise. Conversely, $a \BO{OR}_3^\circ b$ selects $a$ if 
$a \succ_3^\circ b$ and $b$ otherwise.

\begin{figure}[!t]
\centering
\includegraphics[width=4cm]{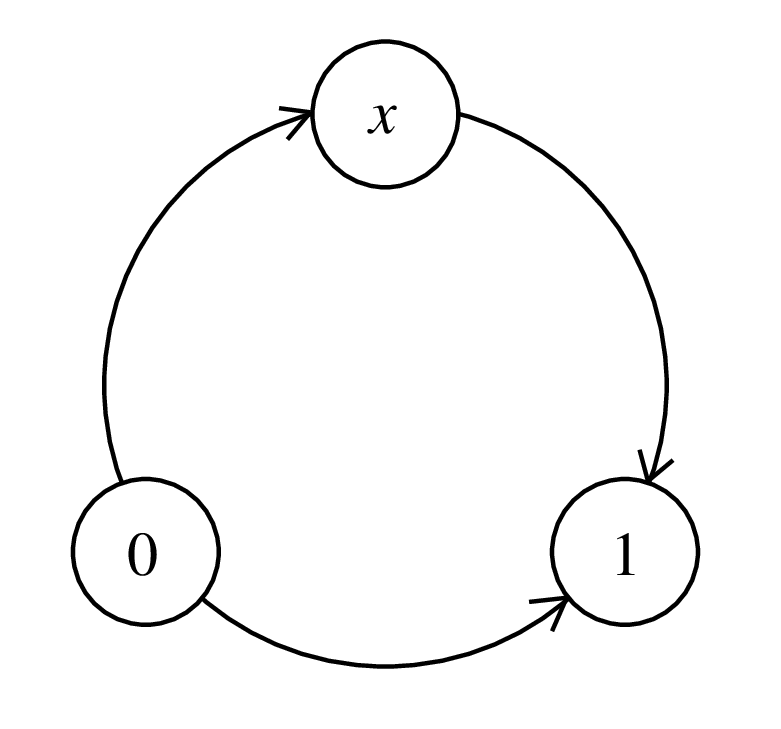}
\caption{Relation (order) for {\L}ukasiewicz logic}
\label{c12x}
\end{figure}

It is similar with definition
of many-valued and real-valued {\L}ukasiewicz logics with AND, OR expressed 
via MIN, MAX respectively \cite{mvlog} that is also relevant to usual Boolean 
logic if to choose notation 0 for {\em false}, 1 for {\em true} and standard 
ordering $0 < 1$. For the case of a three-valued logic with third ``{\em unknown}'' 
value $x$ in such MIN/MAX description the order (i.e., transitive relation) 
$0 \prec_3^{\text{\L}} 1$, $0 \prec_3^{\text{\L}} x$, 
$x \prec_3^{\text{\L}} 1$ should be used, see Fig.~\ref{c12x}.

It corresponds to ordering in {\L}ukasiewicz
three-valued logic and so the notation $1/2$ sometimes is used for the 
third value (denoted here as $x$). But such a logic may not be used 
in construction of the reversible circuit with two ternary ``wires.'' 
It is enough to look on ``truth tables'' for ternary logic \cite{mvlog}
\begin{equation}
\begin{array}{c|ccc}
\BO{AND}_3^{\text{\L}} & 0 & 1 & x \\ \hline
0   & 0 & 0 & 0 \\
1   & 0 & 1 & x \\
x   & 0 & x & x
\end{array}
\qquad
\begin{array}{c|ccc}
\BO{OR}_3^{\text{\L}} & 0 & 1 & x \\ \hline
0  & 0 & 1 & x \\
1  & 1 & 1 & 1 \\
x  & x & 1 & x
\end{array}
\label{luklog}
\end{equation}
Any component of reversible function with two arguments should be ``balanced,''
i.e., each value should be presented an equal number of times like in \Eq{cyclog}.
The property is consequence of possibility to represent a reversible function
as some permutation. 

So for a linear order such a method may not generate balanced 
table, because each element has different number of predecessors
(and successors). But for the cyclic relation depicted on Fig.~\ref{c123} 
a preceding or following element is always unique.  
It could be said, that AND$_3^\circ$, OR$_3^\circ$ are
functions derived from nontransitive (cyclic) arbitration
relation in rock-paper-scissors kind games.

The operations AND$_3^\circ$, OR$_3^\circ$ are not associative 
and not distributive if expressions
contain more than two different values. Yet, 
\[\MO{NOT} (a \BO{AND}_3^\circ b) = (\MO{NOT}\,a) \BO{OR}_3^\circ (\MO{NOT}\,b)\]
is valid for \[\MO{NOT}\,a = (1-a) \bmod 3.\]

It is possible to create the reversible ternary implementation of the binary AND, OR 
using expressions with AND$_3^\circ$, OR$_3^\circ$ defined by \Eq{cyclog}
\[\MO{AND}^\circ \colon (a,b) \mapsto 
\bigl(a \BO{AND}_3^\circ b \, , (b - a) \bmod 3 \bigr),\]
\[\MO{OR}^\circ \colon (a,b) \mapsto 
\bigl(a \BO{OR}_3^\circ b \, , (b - a) \bmod 3 \bigr)\]
\begin{equation}
\begin{array}{|lc|l} 
 \BO{AND}^\circ & & \BO{OR}^\circ \\
\begin{array}{|cc|c|cc|}
 a & b &\rightarrow& a & b \\ \hline
 \mathbf 0 & \mathbf 0 & & \mathbf 0 & 0  \\
 \mathbf 0 & \mathbf 1 & & \mathbf 0 & 1  \\
 0 & 2 & & 2 & 2  \\
 \mathbf 1 & \mathbf 0 & & \mathbf 0 & 2  \\
 \mathbf 1 & \mathbf 1 & & \mathbf 1 & 0  \\
 1 & 2 & & 1 & 1 \\
 2 & 0 & & 2 & 1 \\
 2 & 1 & & 1 & 2 \\
 2 & 2 & & 2 & 0
\end{array}
& &
\begin{array}{|cc|c|cc|}
 a & b &\rightarrow& a & b \\ \hline
 \mathbf 0 & \mathbf 0 & & \mathbf 0 & 0  \\
 \mathbf 0 & \mathbf 1 & & \mathbf 1 & 1  \\
 0 & 2 & & 0 & 2  \\
 \mathbf 1 & \mathbf 0 & & \mathbf 1 & 2  \\
 \mathbf 1 & \mathbf 1 & & \mathbf 1 & 0  \\
 1 & 2 & & 2 & 1 \\
 2 & 0 & & 0 & 1 \\
 2 & 1 & & 2 & 2 \\
 2 & 2 & & 2 & 0
\end{array}
\end{array}
\label{tritri}
\end{equation}

The OR$^\circ$ gate may be also used as a (binary) FANOUT gate, 
if to apply zero to the first input and to use only zero and unit for the second input. 
It may be checked also that the inverse of both gates in \Eq{tritri} may be
used as a (ternary) FANOUT gate for arbitrary value on the first wire 
if to apply zero to the second one. Unlike Toffoli gate which is equivalent
to its own inverse $\rm T^{-1} = T$ presented operations
have longer periods: \( (\MO{AND}^\circ)^{-1} = (\MO{AND}^\circ)^6\),
\( (\MO{OR}^\circ)^{-1} = (\MO{OR}^\circ)^6 \).

The both operations represented in \Eq{tritri} may be performed
with two steps. The first one is $b \mapsto b - a \bmod 3$. The second
step for AND$^\circ$ is controlled subtraction of unit (mod 3) from
$a$ if $b=2$. Contrary, for OR$^\circ$ it is controlled addition of 
unit (mod 3) to $a$ if $b=1$. All such steps are reversible and
resembles the method discussed after \Eq{twotri}.

Let $\mathrm{X}_3 \colon a \mapsto (a + 1) \bmod 3$
is a reversible operation 
with the property $\mathrm X_3^{-1} = \mathrm X_3^2$.
In the quantum computation it may be represented as the matrix
\begin{equation}
 \op{X}_3 = 
 \begin{pmatrix}
  0 & 0 & 1 \\ 1 & 0 & 0 \\ 0 & 1 & 0 
 \end{pmatrix}\!, ~
 \op{X}_3^* = \op{X}_3^{-1}\! = \op{X}_3^2 = 
 \!\begin{pmatrix}
  0 & 1 & 0 \\ 0 & 0 & 1 \\ 1 & 0 & 0 
 \end{pmatrix}\!. 
\label{QX}
\end{equation}

Then in the quantum case the first step discussed above may be expressed 
using $\op{X}_3$ and projectors $\op{P}_k = \ket{k}\bra{k}$  as 
\[ \op{C}_X^* = \op{P}_0 \otimes \op{\Id} + \op{P}_1 \otimes \op{X}_3^*
+ \op{P}_2 \otimes \op{X}_3 \]
and operations used on second step are
\[ \op{C}_2^* = \op{\Id} \otimes (\op{\Id} - \op{P}_2)  + \op{X}_3^* \otimes \op{P}_2, \]
\[ \op{C}_1 = \op{\Id} \otimes (\op{\Id} - \op{P}_1)  + \op{X}_3 \otimes \op{P}_1 \]
respectively. Finally, in such notation it may be written
\[
 \op{\BO{AND}}^\circ =  \op{C}_2^* \op{C}_X^*, \qquad 
 \op{\BO{OR}}^\circ =  \op{C}_1 \op{C}_X^*. 
\]

\section{Note on ``Classical'' Computations on Quantum Computer}

After the early paper of Feynman \cite{FeyComp} 
the question about doing usual computation on  
quantum computer is not widely discussed.
Most attention is devoted to quantum phenomena like 
superposition, entanglement and to quantum algorithms providing
a speedup in comparison with the classical case.

On the other hand, problems of an information processing by quantum systems
may be very actual even for usual algorithms. A necessity for reversible 
operations was already mentioned, but even such gates should
be considered in specific way for quantum systems. As an example may
be mentioned the NOT gate. It simply swaps 0 and 1, but 
for quantum system it might be described by some process started
at some time $t_1$ and finished at $t_2 = t_1+\Delta t$. 

At some moment between $t_1$ and $t_2$ the system may not be described
neither by state $\ket{0}$ nor by state $\ket{1}$ and for the simple
example with an ``ideal'' qubit, i.e., a closed system with only two basic 
states, it may be expressed by some superposition 
$\alpha_t \ket{0} + \beta_t \ket{1}$,  
$|\alpha_t|^2 + |\beta_t|^2 = 1$. So the notion about the specific quantum 
phenomena is reasonable even for the simplest classical algorithm,
if it is implemented by a quantum system.       

\section{Conclusion}

Some unconventional information technologies such as reversible and ternary circuits 
were discussed, which may be relevant to the development of the quantum computing. 
The main theme of the paper is
realization of Boolean functions on a quantum computer using 
qubits and qutrits. Instead of a brute force method of implementation 
of Boolean AND, OR with ternary variables it is
used some variation of a MIN/MAX approach known earlier due to {\L}ukasiewicz 
many-valued logic. This construction is useful in theory of quantum and 
reversible computing, but also may have independent applications.

After finishing initial version of the paper author became aware about
two informal talks \cite{JBN}, where operation AND$_3^\circ$ was
introduced using similar construction with cyclic {\em jan-ken-po}
(the rock-paper-scissors game) relation denoted above as $\prec_3^\circ$.
The cyclic relation is also used in Arrow's Theorem \cite{Arrow}.

%\section*{Acknowledgment}

\newpage
\appendix
\newcommand{\swp}{\bowtie}
\section*{Appendix: ``Completely Cyclic'' Gates}

A gate $G$ {\em respects the cyclic relation} $\prec_3^\circ$, 
if from $G(a,b) = (c,d)$ follows
\[G( a + 1 \bmod 3,b + 1 \bmod 3 ) 
= (c + 1 \bmod 3,d + 1 \bmod 3).\]
The OR$^\circ$, AND$^\circ$ gates do not respect the cyclic 
relation completely, because the second output $ (b - a) \bmod 3$ 
does not have necessary property. 

An example of operation, respecting cyclic relation is
$b \mapsto (2 b - a) \bmod 3$, but  
it is equivalent with  $2 (b + a) \bmod 3$ and the
second output of such gate is symmetric with respect 
to exchange of $a$ and $b$. The first 
output with AND$_3^\circ$, OR$_3^\circ$ 
has the same symmetry and such gate would be irreversible,
because different input pairs 
such as $(a,b)$ and $(b,a)$ are producing  
indistinguishable result.

Both reversibility and cyclic relation can be met, if instead 
of considered modification with $(2 b - a)$ to use such gate as
\[ \MO{OR}^\circ_+ \colon (a,b) \mapsto \bigl( a \BO{OR}_3^\circ b \, \, , 
(a \BO{OR}_3^\circ b  + b - a) \bmod 3 \bigr).\]

The gate OR$^\circ_+$ is reversible, because it can be expressed as
composition of two reversible gates: OR$^\circ$  and 
\[ C_X \colon (a,b) \mapsto \bigl(a, (b + a) \bmod 3 \bigr). \]
The notation $C_X$ is used, because the inverse gate
\[ C_X^{-1} \colon (a,b) \mapsto \bigl(a, (b - a) \bmod 3 \bigr) \]
is classical reversible analogue of quantum gate
$\op{C}_X^{-1} = \op{C}_X^*$ introduced earlier.

The composition of AND$^\circ$ with $C_X$ produces
\[ \MO{AND}^\circ_+ \colon (a,b) \mapsto 
\bigl( a \BO{AND}_3^\circ b \, \, , 
(a \BO{AND}_3^\circ b  + b - a) \bmod 3 \bigr)\]
and compositions of OR$^\circ$, AND$^\circ$
with $C_X^{-1}$ correspond to
\[
\begin{split}
\MO{OR}^\circ_- \colon& (a,b) \mapsto \bigl( a \BO{OR}_3^\circ b \, \, , 
(a \BO{OR}_3^\circ b  + a - b) \bmod 3 \bigr),\\
\MO{AND}^\circ_- \colon& (a,b) \mapsto \bigl( a \BO{AND}_3^\circ b \, \, , 
(a \BO{AND}_3^\circ b  + a - b) \bmod 3 \bigr).
\end{split}
\]

A reversible two-gate is a permutation of pairs
and compact notation with disjoint cycles is used below,
{\em e.g.} tables \Eq{tritri} may be rewritten as
\[
\begin{split}
\MO{AND}^\circ &= (0\,2, 2\,2, 2\,0, 2\,1, 1\,2, 1\,1, 1\,0), \\  
 \MO{OR}^\circ &= (0\,1, 1\,1, 1\,0, 1\,2, 2\,1, 2\,2, 2\,0).  
\end{split} 
\]
In such notation four gates introduced above are expressed as
\begin{equation}
\begin{split}
\MO{AND}^\circ_+ &= (0\,2, 2\,1, 1\,0), \\
 \MO{OR}^\circ_+ &= (0\,1, 1\,2, 2\,0), \\
\MO{AND}^\circ_- &= (0\,1, 0\,2, 2\,0, 2\,1, 1\,2, 1\,0), \\ 
 \MO{OR}^\circ_- &= (0\,1, 1\,0, 1\,2, 2\,1, 2\,0, 0\,2 ).
\end{split} 
\label{allfour}            
\end{equation}   

The compact notation uncovers interesting properties
of the gates, {\em e.g.} 
\[ (\MO{OR}^\circ_-)^{-1} = (\MO{OR}^\circ_-)^5
 = \MO{AND}^\circ_-
\]
and so AND$^\circ_-$ may be constructed from OR$^\circ_-$
and vice versa.
The OR$^\circ_+$, AND$^\circ_+$ change
only three input pairs and may be represented as
\begin{equation}
\begin{split}
\MO{OR}^\circ_+ \colon &(a,b) \mapsto 
\begin{cases}
 (a,b) & a + 1 \neq b \bmod 3, \\
 (a+1 \bmod 3, b + 1 \bmod 3)  & a + 1 = b \bmod 3,
\end{cases}\\
\MO{AND}^\circ_+ \colon &(a,b) \mapsto 
\begin{cases}
 (a,b) & a + 2 \neq b \bmod 3, \\
 (a+2 \bmod 3, b + 2 \bmod 3)  & a + 2 = b \bmod 3.
\end{cases}
\end{split}
\label{twocases}
\end{equation}
The composition of {\em exchange} $\swp \colon (a,b) \to (b,a)$ with  
\Eq{allfour} produces
\[ \swp \MO{AND}^\circ_- = (\MO{OR}^\circ_+)^{-1},
\quad \swp \MO{OR}^\circ_- = (\MO{AND}^\circ_+)^{-1}.\]
So, any gate from \Eq{allfour} is enough for construction
of other gates using repetition of the gate and exchange
of wires.

Let's also consider an operation 
$\MO{N}^\circ_- = (\MO{OR}^\circ_-)^3 =
 (\MO{AND}^\circ_-)^3 $,
\[
\MO{N}^\circ_-  = (0\,1, 2\,1)\, (1\,0, 2\,0)\, (1\,2, 0\,2).
\]
If second input is 2, it works as binary NOT gate for values 0, 1 
on the first input.
Yet another useful expression may be checked directly
\begin{equation}
\MO{N}^\circ_- \colon (a,b) \mapsto (-(a + b) \bmod 3, b).
\label{Ng}
\end{equation}

\begin{figure}[hbt]
\centering
\quad\includegraphics[width=4.5cm]{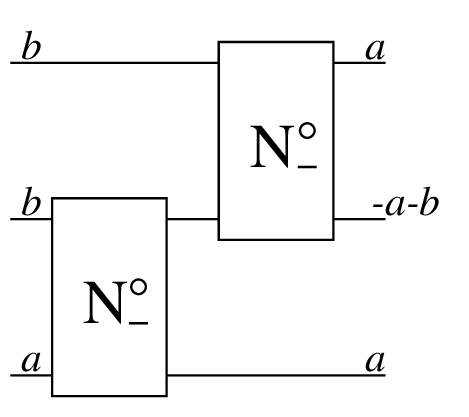}
\caption{FANOUT circuit}
\label{Ngates2}
\end{figure}

Implementation of FANOUT should use at least three wires 
because any composition of gates from \Eq{allfour}
on two wires may not modify pair of equivalent values.
Due to \Eq{Ng} simple construction of FANOUT may
use two N$^\circ_-$ gates, 
Fig.~\ref{Ngates2}. The scheme works for arbitrary $a$ 
and value of $b$ does not matter.

\end{document}